\documentclass[11pt,titlepage]{article}

\usepackage{setspace}
\usepackage{color}

\usepackage{amsmath}

\usepackage{graphicx}

\usepackage[round,numbers,sort&compress]{natbib}
\newcommand{\be}{\begin{equation}}
\newcommand{\ee}{\end{equation}}
\newcommand{\bea}{\begin{eqnarray}}
\newcommand{\eea}{\end{eqnarray}}

\title{Role of ATP-hydrolysis in the dynamics of a single actin filament}

\author{Padinhateeri Ranjith\\
        Physico-Chimie UMR 168\\
    Institut Curie, 26 rue d'Ulm \\
    75248 Paris Cedex 05, France
    \and Kirone Mallick\\
    Service de Physique Th\'{e}orique\\
     CEA Saclay\\
     91191 Gif, France
 \and Jean-Fran\c{c}ois Joanny \\
    Physico-Chimie UMR 168 \\
    Institut Curie, 26 rue d'Ulm \\
    75248 Paris Cedex 05, France
    \and David Lacoste \\
    Laboratoire de Physico-Chimie Th\'{e}orique, \\
    ESPCI, 10 rue Vauquelin, \\
    Paris Cedex 05, France
    }

\date{}

\begin{document}

\maketitle

\abstract{
We study the stochastic dynamics of growth and shrinkage of single
actin filaments taking into account insertion,
removal, and ATP hydrolysis of subunits either according to
the vectorial mechanism or to the random mechanism. In a previous
work, we developed a model for a single actin or microtubule
filament where hydrolysis occurred according to the vectorial
mechanism: the filament could grow only from one end, and was in
contact with a reservoir of monomers. Here we
extend this approach in several ways, by including the dynamics
of both ends and by comparing two possible
mechanisms of ATP hydrolysis. Our
emphasis is mainly on two possible limiting
models for the mechanism of hydrolysis within
a single filament, namely the vectorial or the random model.
We propose a set of experiments to test the
nature of the precise mechanism of hydrolysis within actin
filaments.

\emph{Key words:} Actin; ATP hydrolysis;
Stochastic dynamics}

\clearpage

\section*{Introduction}
Actin monomers polymerize to form long helical filaments, by
addition of monomers at the ends of the filament. The two ends are
structurally different. The addition and removal of subunits at
one end, the barbed end, are substantially faster than at the
other end, the pointed end. In an equilibrium polymer, the
critical concentration at which the on- and off-rates are balanced
must be the same at both ends for thermodynamic reasons
\cite{howard-book}. But actin is not an equilibrium polymer, it is
an ATPase, and ATP is rapidly hydrolyzed after polymerization. Due
to this constant energy consumption, the actin polymer exhibits
many interesting non-equilibrium features; most notably it is able
to maintain different critical concentrations at the two ends
\cite{Fujiwara-Vavylonis-2007}. This allows the existence of a
special steady-state called treadmilling, characterized by a flux
of subunits going through the filament, which has been observed
both with actin as well as with microtubules
filaments~\cite{RPhillips:book}.

The precise molecular mechanism of hydrolysis in actin has
been controversial for many years. For each of the two steps
involved in the hydrolysis (the ATP cleavage and the Pi release),
the possibility of the reaction occurring either at the interface
between neighboring units carrying different nucleotides or at
random location within the filament can be invoked. The vectorial
model corresponds to a limit of infinite cooperativity in which
the hydrolysis of a given monomer depends entirely on the state of
its neighbors, and the random model is a model of zero
cooperativity in which the hydrolysis of a given monomer is
independent of the state of its neighbors. In between these two
limits, models with a finite cooperativity have been considered
\cite{wegner:96,Lipowsky:09}.
A direct evidence for a cooperative mechanism was brought recently
by the authors of Ref.~\cite{Perez_Sci:08}, who observed
GTP-tubulin remnants using a specific antibody. 

Several groups have emphasized the process
of random cleavage followed by random Pi release
\cite{vavylonis:05,Bindschaller_etal}. By studying the
polymerization of actin in the presence of phosphate, the authors
of Ref.~\cite{Fujiwara-Vavylonis-2007} argued that the crucial
step of release of the phosphate is not a simple vectorial process
but is probably cooperative. Since this release of phosphate is
slow, the delay between the completion of hydrolysis and the
polymerization can lead to overshoots which indeed have been
 observed in fluorescence intensity measurements
of pyrene-labelled actin during rapid polymerization as discussed
in \cite{carlsson:bpj:08}. At the single filament level, the
dynamics of depolymerization is also very interesting. The study
of this dynamics provides insights into the underlying mechanism
of hydrolysis in actin as discussed recently in
Refs.~\cite{mitchison:08,Lipowsky:09}.

Although decades of work in the biochemistry of actin have
provided a lot of details on the kinetics of self-assembly of
actin in the absence and in the presence of actin binding
proteins, it is difficult to capture the complexity of this
process without a mathematical model to organize all this
information. To this end, we have studied a non-equilibrium model
for a single actin or microtubule filament \cite{actin-bpj:09}
based on the work of Stukalin et al. \cite{kolomeisky:06}. In this
model, the hydrolysis of subunits inside the filament is a
vectorial process, the filament is in contact with a reservoir of
monomers, and growth occurs only from one end. We have analyzed
the phase diagram of that model with a special emphasis on the
bounded growth phase, and we have discussed some features of the
dynamic instability. Our approach differs from previous
work on the
dynamic instability of microtubules in the following way: the
model is formulated in terms of rates associated with monomer
addition/removal and hydrolysis rather than in terms of
phenomenological parameters such as the switching rates between
states of growth and collapse as done in
Refs.~\cite{Leibler:93,Leibler-cap:96}. This should be a
definite advantage when bridging the gap between the theoretical
model and experiments.

The work of Flyvbjerg et al. \cite{Leibler-cap:96} has
inspired a number of other theoretical models, based on a
microscopic treatment of growth, decay, catastrophe and rescue of
the filament: see in particular Ref.~\cite{Wolynes:06} and
Refs.~\cite{Antal-etal-PRE:07,Antal-etal-JSP:07,sumedha:09}, which
analyze using analytical and numerical methods several aspects of
the dynamic instability of microtubules.

In this paper, we present a model for a single actin
filament which accounts for the insertion, removal, and ATP
hydrolysis of subunits at both ends. It extends our previous work
\cite{actin-bpj:09} in several ways: first by including the
dynamics of both ends and secondly by carrying out simulations for
both mechanisms of hydrolysis -- vectorial and random. In section
\ref{sec: 2 ends} we present the first extension due to the inclusion of both ends, and in section \ref{sec:vectorial_random} we study the two versions of the model for
the hydrolysis within the filament. In the last section
\ref{sec:dynamics}, we examine transient properties of a single
filament using numerical simulations and we show that for these
transient properties, the vectorial and random models lead to
distinct behaviors. This suggests experiments that would allow to
discriminate between the two models.

\section{Vectorial model of hydrolysis with activity at both ends}
\label{sec: 2 ends}

ATP hydrolysis is a two steps process: the first step is
the ATP cleavage which produces ADP-Pi, and which is rapid. The
second step is the release of the phosphate (Pi), which leads to
ADP-actin and which is by comparison much slower~\cite{MFC-bc:86}.
ADP-Pi-actin and ATP-actin have similar critical concentrations but they are
kinetically different species, since they have different on and off
rates as shown in Ref.~\cite{Fujiwara-Vavylonis-2007}.
Nevertheless, from a kinetic point of view, the slow step of release of the
phosphate is the rate-limiting essential step. This suggests that
many kinetic features of actin polymerization can be explained
by a simplified model of hydrolysis, which takes into account only the second
step of hydrolysis and treats actin subunits bound to ATP and actin subunits
bound to ADP-Pi as a single specie.
 This is the assumption of
Ref.~\cite{kolomeisky:06}, which we have used in our published study
\cite{actin-bpj:09} as well as in the present work.
In
other words, what is meant by hydrolysis in all these references
is the step of Pi release. In this section, we assume that this
release of Pi is a vectorial process described as a single
reaction with rate $R$.

Let us recall the main features of the phase diagram of our
previous model which assumes that only one end is growing. The
model has three different phases: two phases of unbounded growth
and one phase of bounded growth. In one phase of unbounded growth
(phase III), both the cap and the bulk of the filament are
unbounded. In this rapidly growing phase, the filament is
essentially made of unhydrolyzed ATP-actin monomers. In the
intermediate phase of
 unbounded growth (phase II), the cap length remains constant
 as a function of time while the length of the filament
  grows linearly with time.
Finally, in the phase of bounded growth (phase I), both cap length
and filament length remain constant on average. This phase is
characterized by a finite average length $\langle l \rangle$ and
by a specific length distribution of the filament which were
calculated in Ref.~\cite{actin-bpj:09}. The phase of unbounded
growth is frequently observed with actin, whereas the intermediate
phase only exists as a steady state in a small interval of
concentration of actin monomers
 near the critical concentration. The intermediate phase can, however, be observed outside
this interval in a transient way, by forcing filaments to
depolymerize through a dilution of the external medium. The phase
of bounded growth of a single filament growing from one end only
has not been observed experimentally so far with actin, but it has
been observed and is well known in microtubules
\cite{Leibler:93,Leibler-cap:96}.

We now extend the single-end model by including dynamics at both
the ends. We keep, as before, the assumption of vectorial
hydrolysis, which means that there is a single interface between
the ATP subunit and ADP subunits, and the assumption of a
reservoir of free ATP subunits in contact with the filament. The
addition of ATP subunits occurs with rate $U$ at the barbed end,
the removal of ATP subunits occurs with rate $W_T^+$ at the barbed
end and with a rate $W_T^-$ at the pointed end. The removal of ADP
subunits occurs at the barbed end only if the cap is zero, with
rate $W_D^+$. At the pointed end, ADP subunits are removed with a
rate $W_D^{-}$. Note that we neglect the possibility of addition
of ATP subunits at the pointed end, this assumption is not
essential but simplifies the analysis.

\begin{figure}
\includegraphics[scale=0.6]{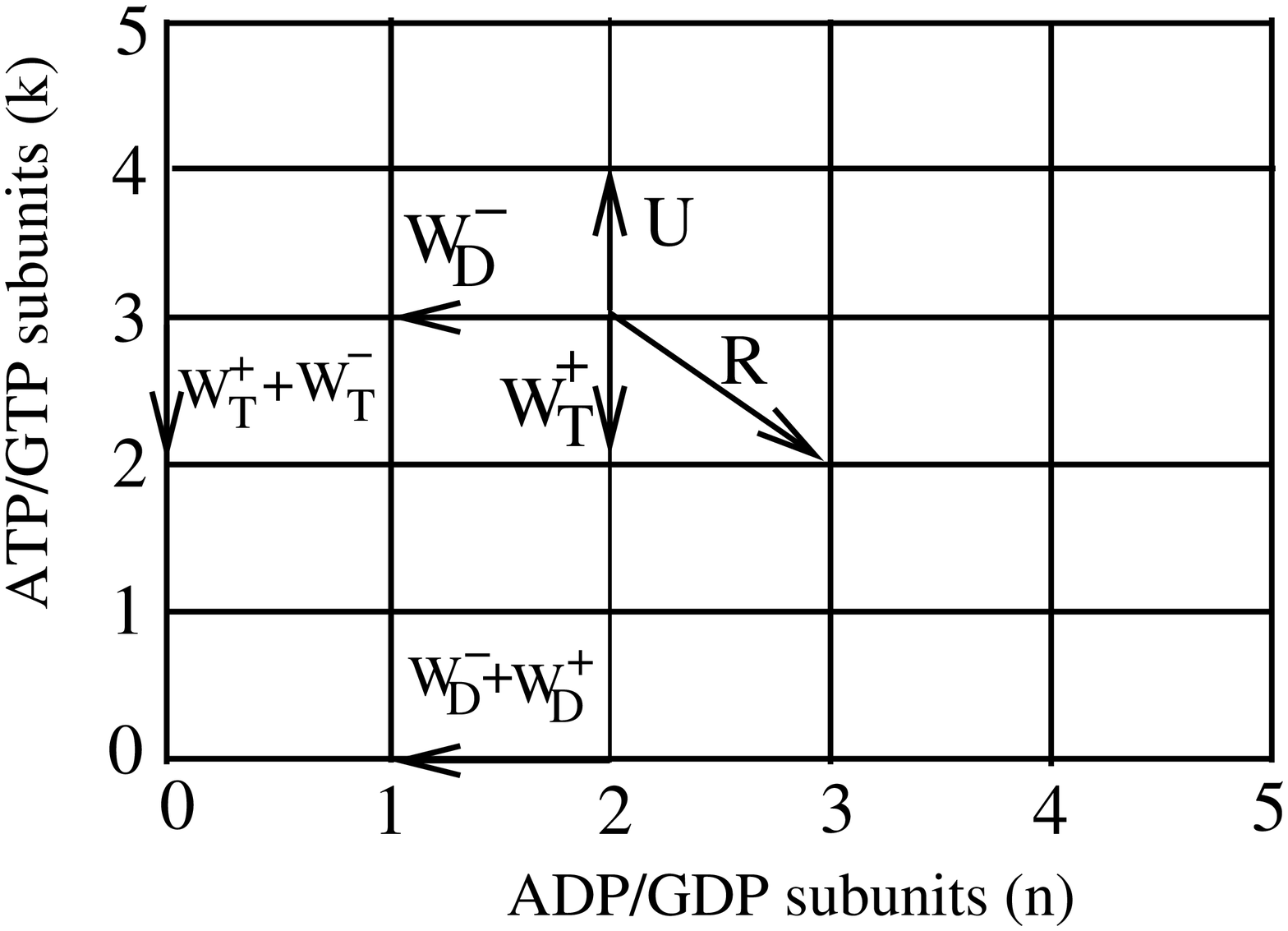}
\caption{\label{fig-model-rates}}
\end{figure}

In Fig. \ref{fig-model-rates},
we have pictorially depicted all these moves discussed above. Furthermore, we
have assumed that all the rates are independent of the
concentration of free ATP subunits $C$ except for the on-rate
which is $U=k_0 C$. All the rates of this model have
been determined precisely experimentally except for $R$. The
values of these rates are given in table \ref{table-rates}.

The state of the filament can be represented in terms of $n$,
the number of ADP subunits and $k$ the number of
ATP subunits. The dynamics of the filament can be mapped onto
that of a random walker in the upper-quarter plane $(n,k)$
with the specific moves as shown in figure \ref{fig-model-rates}.
We find the following steady-state phases (see Appendix A for
details): a phase of bounded growth (phase I), and three phases of
unbounded growth (phase IIA and IIB, phase III). The phase of bounded growth (phases I) and
the phase of unbounded growth with unbounded cap (phase III)
are similar to the corresponding phases in Ref.~\cite{actin-bpj:09}. In the
phase IIA, similar to the phase II of that reference, the filament
is growing linearly in time, with a velocity $v_{IIA}$ but the
average cap length remains constant in time. In the new phase IIB,
the filament is growing linearly in time, with a velocity
$v_{IIB}$ but there is a section of ADP subunits which remains
constant in time near the pointed end (this is analogous to the
cap of ATP subunits near the barbed end in phase IIA).

This phase diagram can be understood from the random walk representation of figure \ref{fig-model-rates}.
The velocity of the random walker in the bulk has components $v_n=(R-W_D^-)d$ along the $x$ axis
 and $v_k=(U-W_T^+-R)d$ along the $y$ axis, where $d$ is the subunit size.  Depending on the signs of these
  quantities, four cases emerge. If $v_n>0$ and $v_k>0$, both the filament and cap length
  increase without bound, this corresponds to phase III. If $v_n<0$ and $v_k<0$, both the filament
  and cap length stay bounded and we have phase I. If $v_n>0$ and $v_k<0$, the cap length remains
  constant in time, but the rest of the filament made of D subunits can be either bounded
  (then we are again in phase I) or unbounded (and we are in phase IIA). Similarly,
  if $v_n<0$ and $v_k>0$, the length of the region of D subunits at the pointed end remain
  constant in time, but the region of T subunits can be either bounded (phase I) or unbounded
   (phase IIB). In phase IIA, the probability
of finding a non-zero cap, \be q=\frac{U}{W_T^+ +R}, \label{def:q} \ee is finite, and the
average filament velocity is (see Appendix A) \be \label{vIIA}
v_{IIA} = [U - W_T^{+} q - W_D^+ (1-q) - W_D^-]d. \ee At the critical
concentration $c=c_A$, $v_{IIA}=0$ and this marks the boundary to
phase I. We find that \be \label{CA} c_A= \frac{(W_D^+ +
W_D^-)}{k_0} \left(\frac{W_T^+ + R}{W_D^+ +R} \right), \ee which
is always larger than the critical concentration of the barbed end
alone. In region III, the velocity is still given by
\be
v_{III}=[U-W_T^+]d.
\label{velocity vIII}
\ee

Similarly, in phase IIB, the probability of finding a
non-zero region of D-subunits $\tilde{q}=R/W_D^-$ is finite, and the
average filament velocity is \be \label{vIIB} v_{IIB}= [U-W_T^+ -
W_T^- (1-\tilde{q}) - W_D^- \tilde{q}]d, \ee which vanishes when
$c=c_B$ at the boundary with phase I, with \be \label{CB}
c_B=\frac{1}{k_0} \left[ (W_T^- - W_D^-) \left( 1-\frac{R}{W_D^-}
\right)+(W_T^++W_D^-) \right]. \ee Note that $W_T^-$ does not
enter in $v_{IIA}$ since the hydrolyzed part of the filament is
always infinitely large in this case, in contrast to the case of
$v_{IIB}$, which depends on both $W_T^-$ and $W_D^-$. Note also that
the velocity $v_{IIA}$ and $v_{IIB}$ are sums of a contribution due
to the barbed end and a contribution due to the pointed end.
This is due to the fact that in all growing phases, the filament is infinitely long
in the steady state, and therefore the dynamics of each end is independent
of the other.

Length fluctuations of the filament are characterized by a
diffusion coefficient which is defined in Appendix A. Since the
dynamics of each end is independent in phase IIA, the diffusion
coefficient of this phase $D_{IIA}$ is the sum of a contribution
from the barbed end and another from the pointed end. From
Ref.\cite{actin-bpj:09} we obtain, \be
D_{IIA}=\frac{d^2}{2}\left[U + W_{T}^+ q + W_{D}^+ (1-q) +
\frac{2(W_D^+-W_T^+)(U+W_D^+ q)}{W_T^+ +R} + W_D^{-}\right].
\label{Delq} \ee where $(W_D^{-}d^2) /2$ is contribution of the
diffusion coefficient due to the pointed end.

\begin{figure}
\includegraphics[scale=0.7]{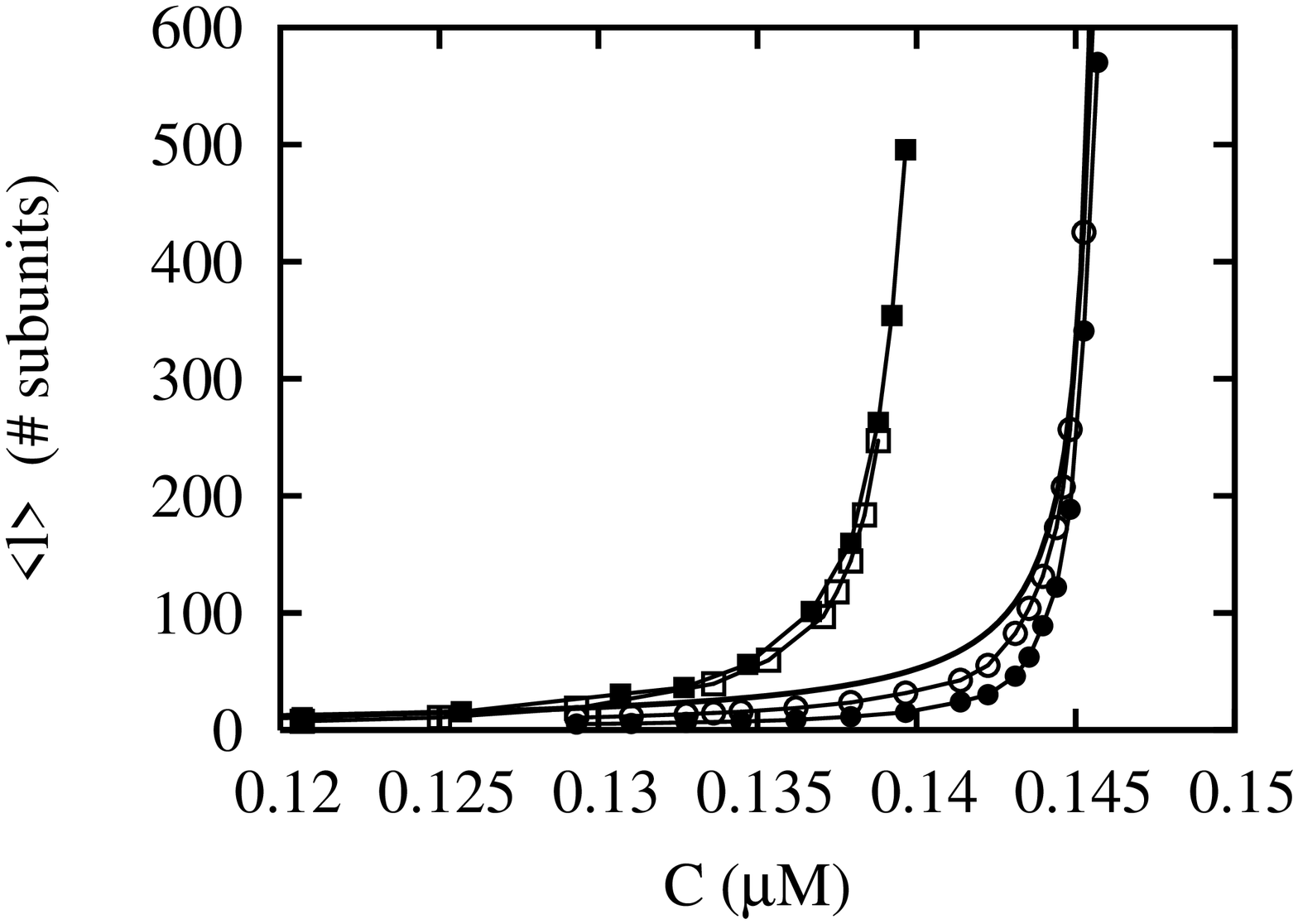}
\caption{\label{fig-avl-bothends} }
\end{figure}

On the boundary lines $c=c_A$ and $c=c_B$, the average filament
velocity vanishes. At this point, the addition of subunits at the
barbed end exactly compensates the loss of subunits at the pointed
end. Such a state is well known in the literature as treadmilling
\cite{hill-book1}. There, the length diverges as
$-D_{IIA}/v_{IIA}$ near $c=c_A$ and similarly as
$-D_{IIB}/v_{IIB}$ near $c=c_B$ as shown in figure
\ref{fig-avl-bothends}, where $D_{IIA}$ and $D_{IIB}$ are
diffusion coefficient in phases $IIA$ and $IIB$. That divergence
is a consequence of the assumption that the filament is in contact
with a reservoir of monomers, in experimental conditions the
maximum length is fixed by the total amount of monomers. In the
bulk of phase I, the average velocity is zero due to a succession
of collapses and nucleations of a new filament. In this phase,
there is a steady state with a well-defined treadmilling average
length.

As mentioned above, since the two ends are far from each other in
the growing phases, they can be treated independently. In the
phase of bounded growth (phase I) however, where the filament
length reaches zero occasionally, the two ends are interacting
strongly. For this reason, a precise description of the phase of
bounded growth is more difficult (see Appendix A). Because of
this, we have computed numerically the average length in
Fig.~\ref{fig-avl-bothends} as function of the free monomer
concentration. In this figure, we compare the case of the filament
with two ends to the case with one end only. We see that there is
a small increase in the critical concentration where the
length diverges and a corresponding lowering of the average
length due to the inclusion of both ends in the model. This effect
is correctly captured by Eqs.~\ref{CA}-\ref{CB}. Note that
although there are large length fluctuations in phase I, the
diffusion coefficient $D_I$ as defined in appendix A is zero in
phase I, because these fluctuations do not depend on time.

\section{Hydrolysis within the filament: a vectorial or random process ?}
\label{sec:vectorial_random}

\subsection{Growth velocity}
As explained earlier, we have used a simplified model for
hydrolysis \cite{kolomeisky:06}, in which the first step of
hydrolysis is absent. The only remaining step, the phosphate
release, is assumed to be a vectorial process. In the following,
we keep this assumption, but we compare the two limiting
mechanisms for the phosphate release, namely the vectorial and the
random processes. All the rates have the same meaning for both
models, except for the hydrolysis rate which is denoted $R$ in the
vectorial model and $r$ in the random model.

We have compared experimental data from \cite{Carlier-etal:86}
together with the two theoretical models, vectorial and random.
Both models successfully account for the observed sharp
bend in the velocity versus concentration plots observed near the
critical concentration as shown in figure \ref{fig:velocity plots with MFC
data}. Below the critical concentration, the velocity is negative for depolymerizing
filaments and it is the velocity of phase II, since phase II extends transiently
below the critical concentration.

Note that the velocities of both models
superimpose, which means that bulk velocity measurements
do not allow to discriminate between these models.
Irrespective of the actual hydrolysis (phosphate release)
mechanism, a fit of this data provides a bound on the value of the
hydrolysis rate in the vectorial model $R$ which is not accurately
determined experimentally. This parameter, was roughly estimated
in Ref.\cite{kolomeisky:06} to be $0.3 s^{-1}$ based on
measurements of Pi release by Melki et al. \cite{Carlier_BC:96}.
The measured hydrolysis rate was multiplied by a typical length to
get the estimate for $R$. Our fit of the data of
Ref.\cite{hill:85}, gives $R=0.1 \pm 0.12 s^{-1}$. This is the
value which we have used for later comparison.

\begin{figure}
\rotatebox{-90}{\includegraphics[scale=0.7]{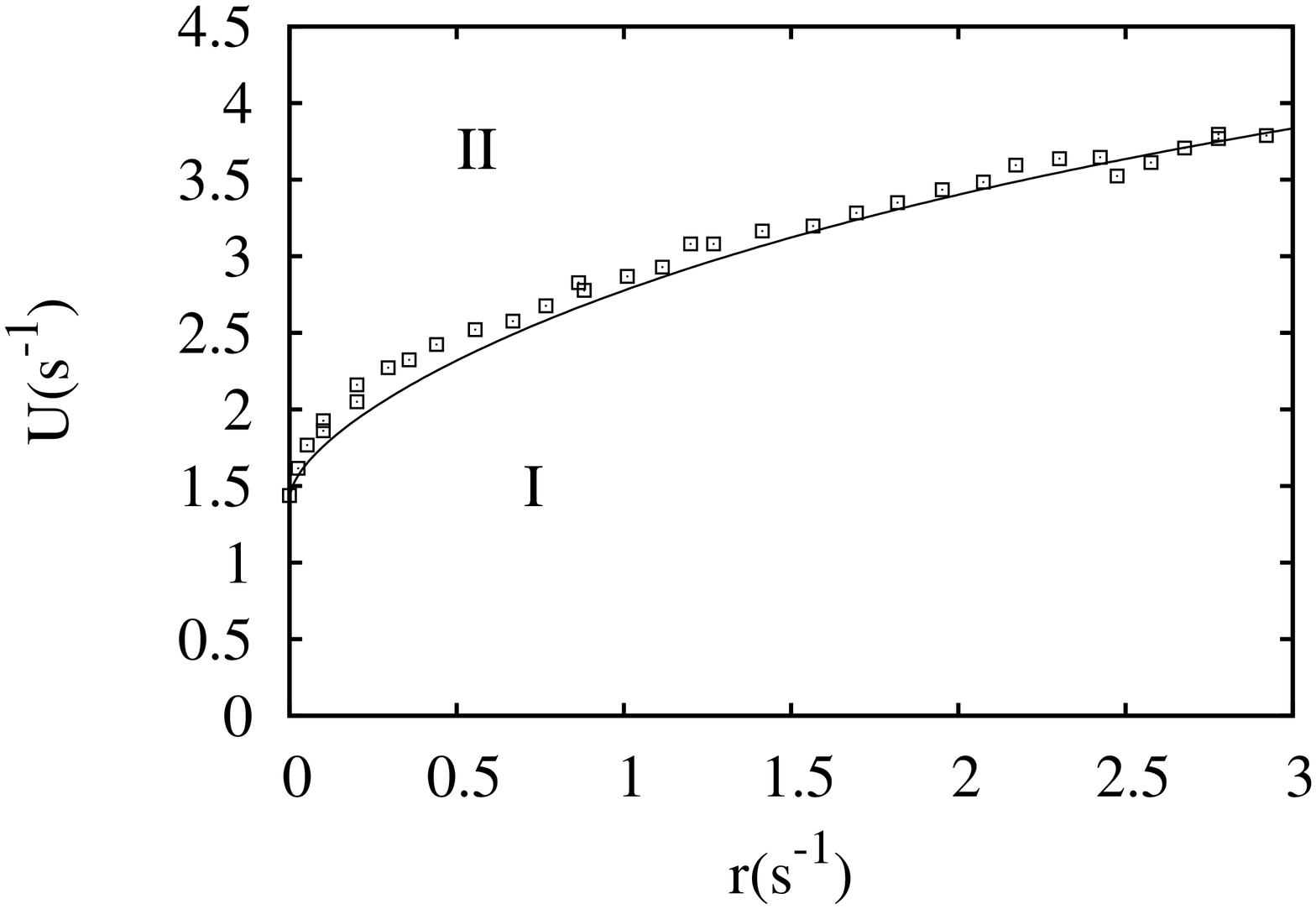}}
\caption{ \label{fig:phases}}
\end{figure}

In figure \ref{fig:phases}, the phase diagram of the random
hydrolysis model is shown. This phase diagram has only two phases
in contrast to the vectorial case, because it can be shown
that the average of the total amount of ATP
subunits $\langle k \rangle$ is always bounded in the random
model. Thus phase III is absent in the random model. In appendix B, we present
details about the derivation of the mean-field equations for the
random model \cite{kolomeisky_BJ:01,wegner:86}. An analytical
expression for the phase boundary between phase I and II is
obtained, which corresponds to the solid line in figure
\ref{fig:phases} and which agrees well with the Monte
Carlo simulations.

\subsection{Length diffusivity}
Length fluctuations are quantified by the length
diffusivity also called diffusion coefficient $D$ which is defined
in Eq.~\ref{ddefn0} of Appendix A. The length diffusivity of
single filaments has been measured using TIRF microscopy by two
groups \cite{fujiwara:02,kuhn-pollard:05}. Both groups reported
rather high values, of the order of 30 monomer$^2$/s. This value
is high when thinking in terms of the rates of assembly and
disassembly measured in bulk \cite{pollard:86,pollard:00}. From
such bulk measurements, one could have expected a length
diffusivity at the critical concentration of 1 monomer$^2$/s; so
an order of magnitude smaller than observed in single filament
experiments.

Several studies have been carried out to explain this discrepancy:
Vavylonis et al.~\cite{vavylonis:05} computed the diffusion
coefficient $D$ as a function of ATP monomer concentration and
found that $D$ is peaked just below the critical concentration and
its maximum is comparable to the value observed in
experiments($\approx 30$ monomer$^2$/s). Stukalin et
al~\cite{kolomeisky:06} obtained from an analytical model the same
large values for $D$ ($\approx 30$ monomer$^2$/s) just above the
critical concentration. Recently, Fass et al. studied the length
diffusivity numerically taking into account filament fragmentation
and annealing, within the vectorial model~\cite{Fass-etal:08}.
They found that high length diffusivity at the critical
concentration cannot be explained by fragmentation and annealing
events unless both fragmentation and annealing rates are much
greater than previously thought. In the limit where their
fragmentation rate goes to zero, they recover the results of
Ref.~\cite{vavylonis:05}.
Others have expressed the opinion that the discrepancy in diffusivity
may be related to experimental errors in the length of the filament
due to out of plane bending of the filaments (M. F. Carlier, private communication).


According to Stukalin et al~ \cite{kolomeisky:06} and Vavylonis et
al \cite{vavylonis:05}, the large length diffusivity observed in
experiments results from dynamic instability-like fluctuations of
the cap. It is important to point out that both papers make very
different assumptions: the first one describes hydrolysis as a
single step corresponding to Pi release with the vectorial
mechanism, whereas the second one describes both steps as random
processes.

\begin{figure}
\includegraphics[scale=0.7]{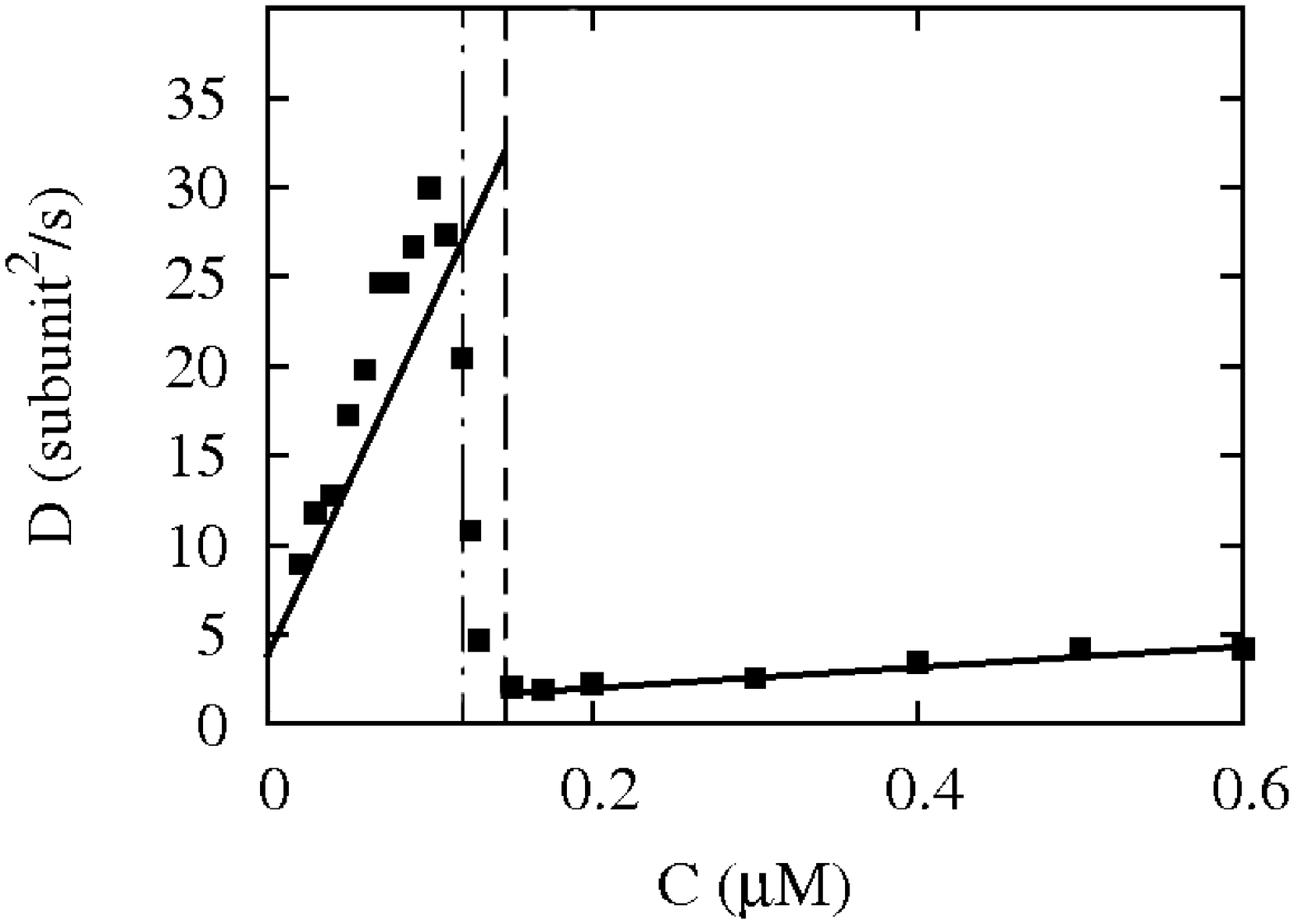}
\caption{\label{cdiff}}
\end{figure}

We have shown in figure \ref{cdiff} the concentration dependance of $D$ for
the vectorial model using analytical expressions provided in the
appendix and similar to that of \cite{kolomeisky:06,actin-bpj:09}.
In this figure, the critical concentration defined as the boundary between
phases I and II almost coincides with the concentration at the boundary between phases II and III,
both are of the order of 0.14 $\mu$M. Above this value, $D$ is indeed small,
the expected estimate of 1 monomer$^2$/s is indeed recovered there
because the contribution of hydrolysis is negligible.
Near the critical concentration, however, the fluctuations are much
larger, for a reason which is similar to the reason that leads to
large fluctuations near critical points in condensed matter
systems \cite{tom}. Here, hydrolysis which is known to destabilize
filaments, has a larger effect. It
leads to large fluctuations of the cap, and ultimately to a large
length diffusivity.
Note that the region below the critical concentration
corresponds to the transient extension of phase II discussed in the previous
section. If the fluctuations were probed there for a very long time, one would find $D=D_I=0$,
characteristic of phase I.

In figure \ref{cdiff}, we have compared these analytical
results obtained for the vectorial model
with numerical results obtained for the random model.
In the random model, we use Monte Carlo
simulations to calculate a time dependent diffusion coefficient
$D(t)$, defined as $D(t)= \frac{1}{2} \frac{d}{dt}
\left(\langle l^2 \rangle -\langle l
\rangle^2 \right)$. For concentrations larger than the critical
concentration, the initial condition is $l
(t=0)=0$, whereas for concentrations smaller than the critical concentration, the
initial condition was a very long filament ($l (t=0) >10^6$
subunits) with all subunits in the hydrolyzed state. On a large time window,
we find that $D(t)$ is approximately time independent, and we interpret that value as
the length diffusivity of the random model.
Our results fully agree with that of
Ref.~\cite{vavylonis:05}, and with that of Ref.~\cite{Fass-etal:08}
in the limit of zero fragmentation rate. The length diffusivity
indeed reaches a maximum of the order of 30 monomer$^2$/s below
the critical concentration. As shown in that figure, there is only
a small difference of length diffusivity in the vectorial case as
compared to the random case: the maximum of the curve for the
random model occurs at a smaller concentration than in the
vectorial model. The fact that we are able to reproduce a similar
curve as in Refs.~\cite{vavylonis:05,Fass-etal:08} justifies our
simplifying assumption of describing the hydrolysis as a single
step associated with the release of phosphate rather than taking
into account the two steps as done in these references. More
importantly it confirms the idea that the length diffusivity of
actin, near critical concentration, is dominated by a process similar to the dynamic instability, which
is essentially captured by the vectorial model.

To make further progress, it would be very useful to reproduce
experiments similar to those of Ref.~\cite{fujiwara:02}, on single
filaments for various monomer concentrations, to confirm the
scenario presented above for the length fluctuations of actin.
Given that the predictions of the random and vectorial model are
rather close to each other as shown in our figure \ref{cdiff}, it
is likely that it will be difficult to distinguish between these
models from measurements of the concentration dependence of the
length diffusivity. One reason for which the length diffusivity of
the two models are very close to each other is that a very small
value of the hydrolysis rate $r$ (as estimated from experiments)
has been used, we have observed that if this parameter had a
larger value than expected, the predictions of the vectorial and
random model would differ much more.

\section{Dynamics of the filament in transient regimes}
\label{sec:dynamics} Since it appears difficult to distinguish the
vectorial from the random model using measurements
of growth velocity or length diffusivity, one can turn
to an analysis of the dynamics of the filament length in polymerization \cite{fujiwara:02}
or in depolymerization \cite{mitchison:08,Lipowsky:09} to
discriminate between the two models. Here, we focus on the
dynamics of polymerization of a single filament, in the presence
of a constraint of conservation of the total number of subunits
(free+polymerized). This constraint leads to a steady state with a
constant average length for the filament. We compare the time it
takes for the filament to be fully hydrolyzed to the time that it
takes to reach the steady state length. We also discuss the
corresponding length fluctuations as a function of time. We argue
that both measurements (the lag time of hydrolysis and the time
resolved fluctuations) can distinguish between the two mechanisms of
hydrolysis.

In Fig.~\ref{Lt-fixedC1} we show the filament length as well as
its variance, as a function of time, for both vectorial and random
models. Using Monte Carlo simulations we computed $l(t)$, starting
from $l(t=0)=0$, for 1000 different realizations and calculated
$\sigma^2 (t)=\langle l^2 \rangle -\langle l \rangle^2$.
Concerning the lag time of hydrolysis, we have observed that in
simulations of the vectorial model, the filament typically reaches
its steady state length long before it has been completely
hydrolyzed. The time when this happens corresponds to the point
where the two curves meet in Fig.~\ref{Lt-fixedC1} (a). This
characteristic time since only one end is involved, is $t_H \simeq
\langle l \rangle/R$ where $R$ is the hydrolysis rate in the
vectorial model \cite{actin-bpj:09}. From the figure we find that
$t_H \simeq 3500/0.3 \simeq 11000$s $\simeq 180$ min (with $R
\simeq 0.3$, which is much longer than the typical time to reach
the steady state $t_{SS} \simeq  \langle l \rangle/v \simeq
3500/(11.6 \times 0.7 -1.4) = 520$s. In contrast to this, in the
random model, the time for completion of hydrolysis is comparable
to the time to reach steady state (see Fig.~\ref{Lt-fixedC1} (b))
as both the filament and the ADP part have similar growth
dynamics.

In practice, this lag time of hydrolysis may be difficult to
measure on single filaments since the ATP subunits and ADP
subunits can not be distinguished easily experimentally. In view
of the previous section, on the role of ATP hydrolysis in length
diffusivity, we suggest to study instead the length fluctuations
of the filament as a function of time. Such a quantity is
accessible from image analysis of single filaments with TIRF for
instance. We have simulated the variance of the length
fluctuations $\sigma(t)^2=\langle l(t)^2 \rangle- \langle l(t)
\rangle^2$ as function of time, for the vectorial model and random
model, as shown in Fig.~\ref{Lt-fixedC1} (c) and (d) respectively.
At early times, this variance is linear in time, and the slope
corresponds to the length diffusivity discussed in previous
section, because the constraint of conservation of monomers plays
no role at short times. Once the steady state has been reached, we
find that the variance of the vectorial model shows a sharp
increase when $t \ge t_H$, while the variance of the random model
shows no significant change. The approximately constant variance
of the random model is intermediate between the variance of the
vectorial model before and after the jump.

Thus contrary to velocity and length diffusivity measurements, an
analysis of either the lag time of hydrolysis or of the time
dependence of the length fluctuations provide a direct signature
of the underlying mechanism of hydrolysis.

\section{Conclusion}
In this article, we have analyzed several aspects of the
dynamics of a single actin filament.
Many results discussed above could be extended
\emph{mutatis mutandis} to the case of microtubules.

We have constructed a phase diagram, which summarizes all the
possible dynamical phases of an actin filament with two active
ends and vectorial hydrolysis in its inside. We have found that
quantities like the filament velocity and the length diffusivity
show similar behavior for both vectorial and random model of
hydrolysis. We propose that measuring the length fluctuations of a
single filament as a function of time can distinguish between the
two models for hydrolysis (or to be more precise to the step of
phosphate release). Although more experimental and theoretical
work are needed, studies of the dynamics of the length of single
filaments during polymerization \cite{fujiwara:02} and during
depolymerization \cite{mitchison:08,Lipowsky:09} suggest a
mechanism of phosphate release which is not purely vectorial or
purely random, but rather partially cooperative.

We hope that our study will contribute to the understanding of the
non-equilibrium self-assembly of actin/microtubule filaments.

 \subsection{Acknowledgements}
We thank M. F. Carlier, J. Baudry, I. Fujiwara and A. Kolomeisky
for illuminating discussions. We also acknowledge discussions with
S. Sumedha, B. Chakraborty and F. Perez. We thank D. Blair for his
contribution to the numerical study of the random model. We
acknowledge support from the Indo-French Center CEFIPRA (grant No.
3504-2), and from Chaire Joliot (ESPCI).

\newpage
\appendix
\noindent {\bf {\large Appendix}}
\section{Equations of the vectorial model with two ends}
Let $P(n,k,t)$ be the probability of having $n$ hydrolyzed ADP
subunits  and  $k$ unhydrolyzed ATP subunits at time $t$, such
that $l=(n+k)d$ is the total length of the filament.  It obeys the
following master equation: For $k>0$ and $n > 0$ we have \bea
\frac{dP(n,k,t)}{dt}&=&U P(n,k-1,t)+W_T^{+} P(n,k+1,t)+R
P(n-1,k+1,t) \nonumber \\
&&+W_D^{-} P(n+1,k)-(U+W_T^{+} +R+W_D^{-})P(n,k,t). \label{pnk}
\eea For $k>0$ and $n =0$ \bea
\frac{dP(0,k,t)}{dt}&=&U P(0,k-1,t)+(W_T^{+}+W_T^{-}) P(0,k+1,t) \nonumber \\
&&+W_D^{-} P(1,k)-(U+W_T^{+} + W_T^{-} +R)P(0,k,t). \label{pnk01}
\eea For $k=0$ and $ n \ge 1$ we have, \be
\frac{dP(n,0,t)}{dt}=(W_D^{+}+W_D^{-}) P(n+1,0,t)+W_T^{+}
P(n,1,t)+R P(n-1,1,t)-(U+W_D^{+}+W_D^{-})P(n,0,t). \label{pn0} \ee
If $k=0$ and $n=0$, we have \be \frac{dP(0,0,t)}{dt}=
(W_T^{+}+W_T^{-}) P(0,1,t)+(W_D^{+} + W_D^{-})P(1,0,t) -
UP(0,0,t). \label{p00} \ee

We define the following generating functions \bea G(x, y,t) &=
&\sum_{n \ge 0} \sum_{k \ge 0} P(n,k,t) x^n y^k,  \\  \label{Glg}
F_k(x,t) &= & \sum_{n \ge 0} P(n,k,t) x^n, \label{fkdefine} \\
H_n(y,t) &= &\sum_{k \ge 0} P(n,k,t) y^k. \eea Normalization
imposes that at all times $t$, \be G(1,1,t)=\sum_{n=0}^{\infty}
\sum_{k=0}^{\infty} P(n,k,t)=1. \ee

Using eqs \ref{pnk}, \ref{pnk01}, \ref{pn0} and \ref{p00}, we
obtain the evolution equation for $G(x,y,t)$ \noindent \bea
\frac{dG(x, y, t)}{dt}&=& \left[ U\left(y-1\right) + W_T^{+}
\left(\frac{1}{y}-1\right)
+ R \left(\frac{x}{y}-1\right)+ W_D^{-}\left(\frac{1}{x}-1 \right) \right]G(x, y,t) \nonumber \\
&-& \left[ W_T^{+} \left(\frac{1}{y}-1\right) + R
 \left(\frac{x}{y}-1\right)+W_D^{+}\left(1-\frac{1}{x}\right)
 \right]F_0(x,t) \nonumber\\
 &-& \left[ W_D^{-} \left(\frac{1}{x}-1\right) +W_T^{-}\left(1-\frac{1}{y}\right)
 \right]H_0(y,t) \nonumber\\
  &-& \left[ W_D^{+} \left(\frac{1}{x}-1\right) -W_T^{-}\left(1-\frac{1}{y}\right)
 \right]P(0,0,t).
 \label{dgdt}
\eea

From $G(x,y,t)$, the following quantities can be obtained: the
velocity of the filament, which is \be v =\lim_{t \to
\infty}\frac{ d \langle l \rangle}{dt}=d \lim_{t \to \infty}
\frac{\partial }{\partial
x}\left(\frac{dG(x,x,t)}{dt}\right)_{x=1} \label{vdefn0}, \ee the
diffusion coefficient characterizing filament length fluctuations
\bea
D&= &\lim_{t \to \infty} \frac{1}{2} \frac{d}{dt}\left(\langle l^2 \rangle-\langle l \rangle^2\right) \nonumber\\
&=&d^2 \lim_{t \to \infty} \left[\frac{1}{2}\frac{\partial^2
}{\partial x^2}\left(\frac{dG(x,x,t)}{dt}\right)
+\frac{1}{2}\frac{\partial }{\partial
x}\left(\frac{dG(x,x,t)}{dt}\right) \right. \nonumber\\&&
\left.-\left(\frac{\partial G(x,x,t) }{\partial
x}\right)\frac{\partial }{\partial
x}\left(\frac{dG(x,x,t)}{dt}\right) \right]_{x=1}. \label{ddefn0}
\eea
The average cap velocity is \be J=d \lim_{t \to \infty} \frac{ d
\langle k \rangle}{dt}=d \lim_{t \to \infty} \frac{\partial
}{\partial y}\left(\frac{dG(1,y,t)}{dt}\right)_{y=1}
\label{Jdefn0}, \ee and the diffusion coefficient characterizing
the fluctuations of the cap is

 \bea D_c&=&
d^2 \lim_{t \to \infty}\frac{1}{2}
\frac{d}{dt}\left(\langle k^2 \rangle-\langle k \rangle^2\right) \nonumber\\
 &=&d^2 \lim_{t \to \infty} \left[\frac{1}{2}\frac{\partial^2 }{\partial
y^2}\left(\frac{dG(1,y,t)}{dt}\right) +\frac{1}{2}\frac{\partial
}{\partial
y}\left(\frac{dG(1,y,t)}{dt}\right) \right. \nonumber \\
& &\left.- \left(\frac{\partial G(1,y,t) }{\partial
y}\right)\frac{\partial }{\partial
y}\left(\frac{dG(1,y,t)}{dt}\right) \right]_{y=1}. \label{dcdefn0}
 \eea

\subsection*{Phase diagram and average length in the bounded phase}
To construct the phase diagram, we first focus on steady-states
solutions of Eq.~\ref{dgdt}, which are such that $dG(x,y,t)/dt=0$.
The obtained equation for $G(x,y)$ involves the following time
independent quantities \bea
F_0(x) & = & G(x,0)  =  \sum_{n \ge 0} P(n,0) x^n,  \\
H_0(y) & = & G(0,y)= \sum_{k \ge 0} P(0,k) y^k,  \\
P(0,0) & = & F_0(0) = H_0(0) = G(0,0), \eea which are coupled back
to $G(x,y)$.

Progress can be made by considering two particular cases for $x=1$
and $y=1$ of this expression for $G(x,y)$. This leads to
\bea
R-W_T^- & = & F_0(1) \left( R+W_D^+ \right) - W_D^- H_0(1) -
P(0,0) W_T^+, \label{particular cases G_1} \\
U-R-W_T^+ & = & - F_0(1) \left( R+W_T^+ \right) + W_T^- H_0(1) -
P(0,0) W_T^-. \label{particular cases G_2} \eea
These two
equations involve three unknowns $F_0(1)$: the probability that
the cap is zero, $H_0(1)$: the probability that the $D$ part of
the filament is zero, and $P(0,0)$: the probability that the
filament is in the state of monomers. Note that $P(0,0)=0$ in
phases of unbounded growth whereas $P(0,0)>0$ in the phase of
bounded growth.

In the random walk representation of figure \ref{fig-model-rates},
 the velocity of the random walker in the bulk has components $v_n=(R-W_D^-)d$ along the $x$ axis
 and $v_k=(U-W_T^+-R)d$ along the $y$ axis. Depending on the signs of these
  quantities, four cases emerge. If $v_n>0$ and $v_k>0$, both the filament and cap length
  increase without bound (phase III) which means that $F_0(1)=H_0(1)=P(0,0)=0$.
  If $v_n<0$ and $v_k<0$, both the filament
  and cap length stay bounded (phase I) and $F_0(1)>0, H_0(1)>0$ and $P(0,0)>0$.

If $v_n>0$ and $v_k<0$, the cap length remains
  constant in time which means $F_0(1)>0$, but the rest of the filament made of D subunits
can be either bounded (for $H_0(1)=P(0,0)=0$, which corresponds to
phase I) or unbounded (for $H_0(1)=P(0,0)>0$ which corresponds to phase
IIA). When reporting the condition $H_0(1)=P(0,0)=0$ into
Eqs.~\ref{particular cases G_1}-\ref{particular cases G_2} and
solving for $F_0(1)$, one finds that the phase of bounded growth
occurs when $U/(R+W_T^+) < (W_D^+ + W_D^-)/(R+W_D^+)$, and the
boundary to the phase of unbounded growth corresponds to replacing
the unequal sign by an equal sign.

An alternative way to find this condition is to start from the
time dependent evolution equation of $G(x,y,t)$ of Eq.~\ref{dgdt}
and impose $H_0(y,t)=P(0,0,t)=0$. We end up with two coupled
dynamical equations for $F_0(x,t)$ and $G(x,y,t)$. The way to
obtain the velocity and diffusion coefficient in phase IIA from
these equations is explained in details in the appendix of
 Ref.~\cite{actin-bpj:09}. The result is the
expression of $v_{IIA}$ given in Eq.~\ref{vIIA}, and the
expression of $D_{IIA}$ of Eq.~\ref{Delq}. As expected, the
condition that marks the boundary between phase IIA and phase I
corresponds to $v_{IIA}=0$.

  Similarly, if $v_n<0$ and $v_k>0$, the length of the region of D subunits at the pointed end remains
  constant in time, and the region of T subunits can be either bounded (phase I) or unbounded
   (phase IIB). By either method, one obtains the velocity in the phase IIB given in Eq.~\ref{vIIB}, and
the condition that marks the boundary to phase I, which
corresponds to $v_{IIB}=0$.

In Ref.~\cite{actin-bpj:09}, an explicit expression for the
average length in the phase
 of bounded growth was obtained by a method of cancellation of
 poles of $G(x,y)$. Unfortunately,
this method does not allow us to derive the expression of $G(x,y)$ here, because
the rates $W_T^- \neq 0$ and $W_D^- \neq 0$ lead to an additional
unknown $H_0(y,t)$ in Eq.~\ref{dgdt} which makes the problem much
more difficult to solve. For this reason, we
could not derive an explicit expression for the average length in
this case, and we investigated this quantity only numerically.

\section{Mean-field equations of the random model}
\begin{figure}
\includegraphics[scale=0.7]{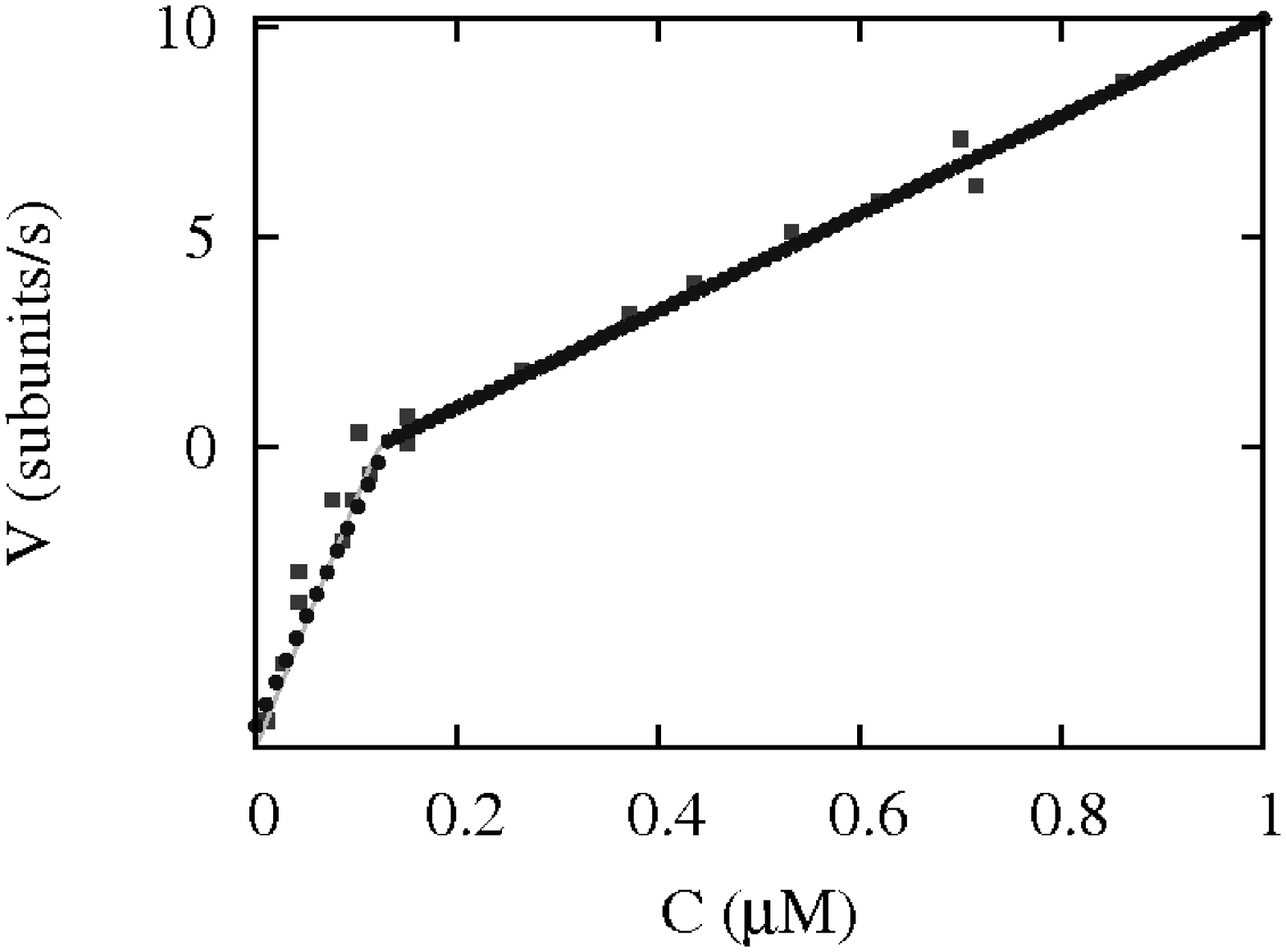}
\caption{\label{fig:velocity plots with MFC data}}
\end{figure}

We explain in this appendix how the velocity of the filament in the random model is obtained
from a mean-field approach. This appendix is provided mainly for pedagogical reasons,
since the solution has already appeared in Ref.~\cite{kolomeisky:06} and Ref.~\cite{wegner:86}.
For simplicity, we focus on the case where growth and shrinking occur only from one end, which we
number as the first site $i=1$.
We use the same notations for the rates as in the vectorial model except for the hydrolysis rate,
which is denoted $r$ in the random model. For a given configuration, we introduce for each subunit
$i$ inside the filament an occupation number $\tau_i$, such that $\tau_i=1$ if the subunit binds ATP
and $\tau_i=0$ otherwise. In the reference frame associated with the end of the filament, the equations
for the average occupation number are
\be \frac{d \langle \tau_1 \rangle}{dt}=U (1- \langle \tau_1 \rangle ) - W_T \langle \tau_1 (1-\tau_2)
\rangle + W_D \langle \tau_2 (1-\tau_1) \rangle - r \langle \tau_1 \rangle, \label{recursion1} \ee

\bea \frac{d \langle \tau_i \rangle}{dt} & = & U ( \langle \tau_{i-1} (1-\tau_i) \rangle -
\langle \tau_{i} (1-\tau_{i-1}) \rangle ) + W_T \langle \tau_1 [ (1-\tau_i) \tau_{i+1} -
\tau_i ( 1- \tau_{i+1} ) ] \rangle \nonumber \\
& + & W_D \langle (1-\tau_1) [ (1-\tau_i) \tau_{i+1} -
\tau_i ( 1- \tau_{i+1} ) ] \rangle - r \langle \tau_i \rangle. \label{recursioni} \eea
In a mean-field approach, the effect of correlations $\langle \tau_i \tau_j \rangle$ are neglected,
\emph{i.e.} these correlations are replaced by $\langle \tau_i \rangle \langle \tau_j \rangle$
(and similarly for averages of product of three occupation numbers).
At steady state, the left-hand sides of Eqs.~\ref{recursion1}-\ref{recursioni} are both zero,
which leads to recursion relations for the $\langle \tau_i \rangle$. Note that $\langle \tau_i
\rangle$ is denoted as $a_i$ in Ref.~\cite{wegner:86} and as $P_i$ in Ref.~\cite{kolomeisky:06}.
We still denote $\langle \tau_1 \rangle=q$, since it represents the probability that the terminal
unit binds ATP. It is the analog of the parameter defined in Eq.~\ref{def:q} for the vectorial model, which is now a more complicated function of the rates.
   The recursion relations have a solution of the form for $i\geq1$,
   \be \frac{ \langle \tau_{i+1} \rangle}{\langle \tau_i \rangle}=
   \frac{U- q (W_T+r)}{U- q W_T}. \label{recursion} \ee
Combining Eqs.~\ref{recursion1}-\ref{recursion}, one obtains the following cubic equation for $q$
\bea
& & (W_T + r)(W_T-W_D) q^3  +  (U W_D - 2 U W_T + W_D W_T + W_D r \\
& - & W_T r - W_T^2) q^2 + U (U- W_D + 2 W_T + r) q - U^2 = 0.
\label{cubic eq}
\eea
This cubic equation has three solutions, but only one solution is such that $0 \leq q \leq 1$. The rate of elongation of the filament can be obtained by reporting that solution into
\be v=\frac{d \langle l \rangle}{dt}= [U - W_T q - W_D (1- q)]d.
\ee
In figure \ref{fig:velocity plots with MFC data}, this velocity $v$ is shown as function of the concentration of free monomers.
For low values of $r$, the velocity of the random and vectorial model are identical, as $r$ is increased the velocity of the random model starts to deviate from the curve of the vectorial model.
By imposing the condition $v=0$, one obtains the phase boundary shown in the solid line in figure \ref{fig:phases}.


\clearpage
\begin{table}
\section*{Tables}
\begin{tabular}{|c|c|c|c|}
\hline
&   &  & References \\
\hline
On rate of T subunits at the barbed end & $k_0$ ($\mu$M$^{-1}s^{-1}$) & 11.6 & \cite{howard-book,kolomeisky:06} \\
\hline
Off-rate of T subunits at the barbed end & $W_T^{+} $($s^{-1}$) &1.4 &\cite{howard-book,kolomeisky:06} \\
\hline
Off-rate of T subunits at the pointed end & $W_T^{-} $($s^{-1}$) &0.8 &\cite{howard-book,kolomeisky:06} \\
\hline
Off-rate of D subunits at the barbed end & $W_D^{+}$($s^{-1}$)&7.2 &\cite{howard-book,kolomeisky:06} \\
\hline
Off-rate of D subunits at the pointed end & $W_D^{-}$($s^{-1}$)&0.27 & \cite{howard-book,kolomeisky:06}\\
\hline
Hydrolysis rate (vectorial model) & $R$ ($s^{-1}$)& 0.1-0.3 &\cite{kolomeisky:06} \\
\hline
Hydrolysis rate (random model) & $r$ (s$^{-1}$)& 0.003 & \cite{kolomeisky:06,vavylonis:05,carlsson:bpj:08}\\
\hline
\end{tabular}
\caption{Various rates used in the model and corresponding references. The conditions are that of a low ionic
strength buffer.
 \label{table-rates}}
\end{table}

\newpage
\section*{Figure Legends}
\subsection*{Figure~\ref{fig-model-rates}} Schematic diagram representing the addition
of subunits with rate $U$, removal with
rates $W_T^+$,$W_T^-$ and $W_D^{+}$, and hydrolysis with rate $R$, which can
only occur at the interface between T and D monomers in the vectorial model. Note that
two new rates $W_T^-$ and $W_D^-$ have been added as compared to Ref.~\cite{actin-bpj:09}.

\begin{figure}
\includegraphics[scale=0.6]{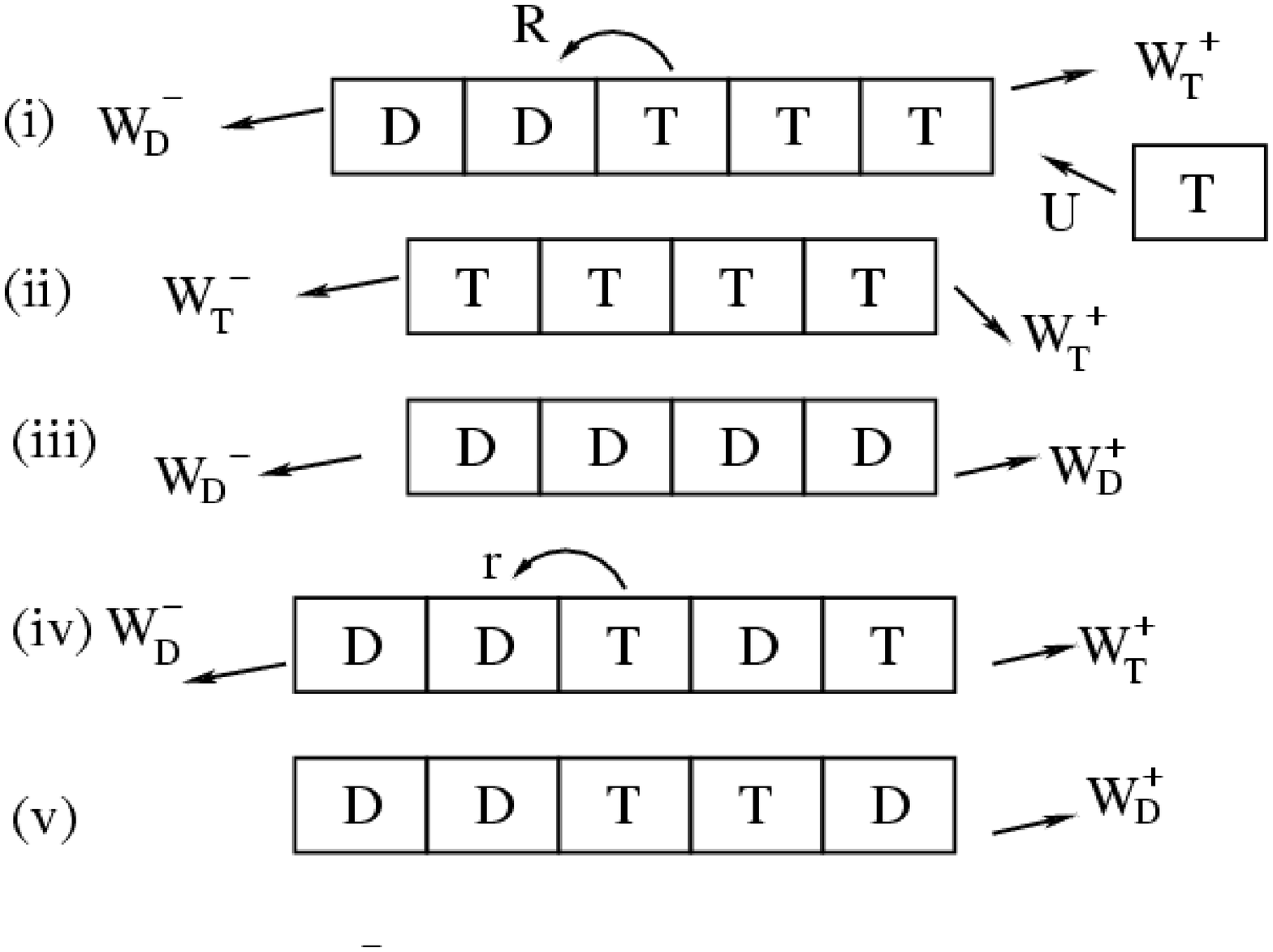}
\caption{\label{walk-2end-bw}}
\end{figure}

\subsection*{Figure~\ref{walk-2end-bw}} Representation of the various possible moves for actin dynamics.
(i), (ii) and (ii) depict different cases for vectorial hydrolysis. (iv) and (v) depict
 cases for random hydrolysis.

\begin{figure}
\includegraphics[scale=0.7]{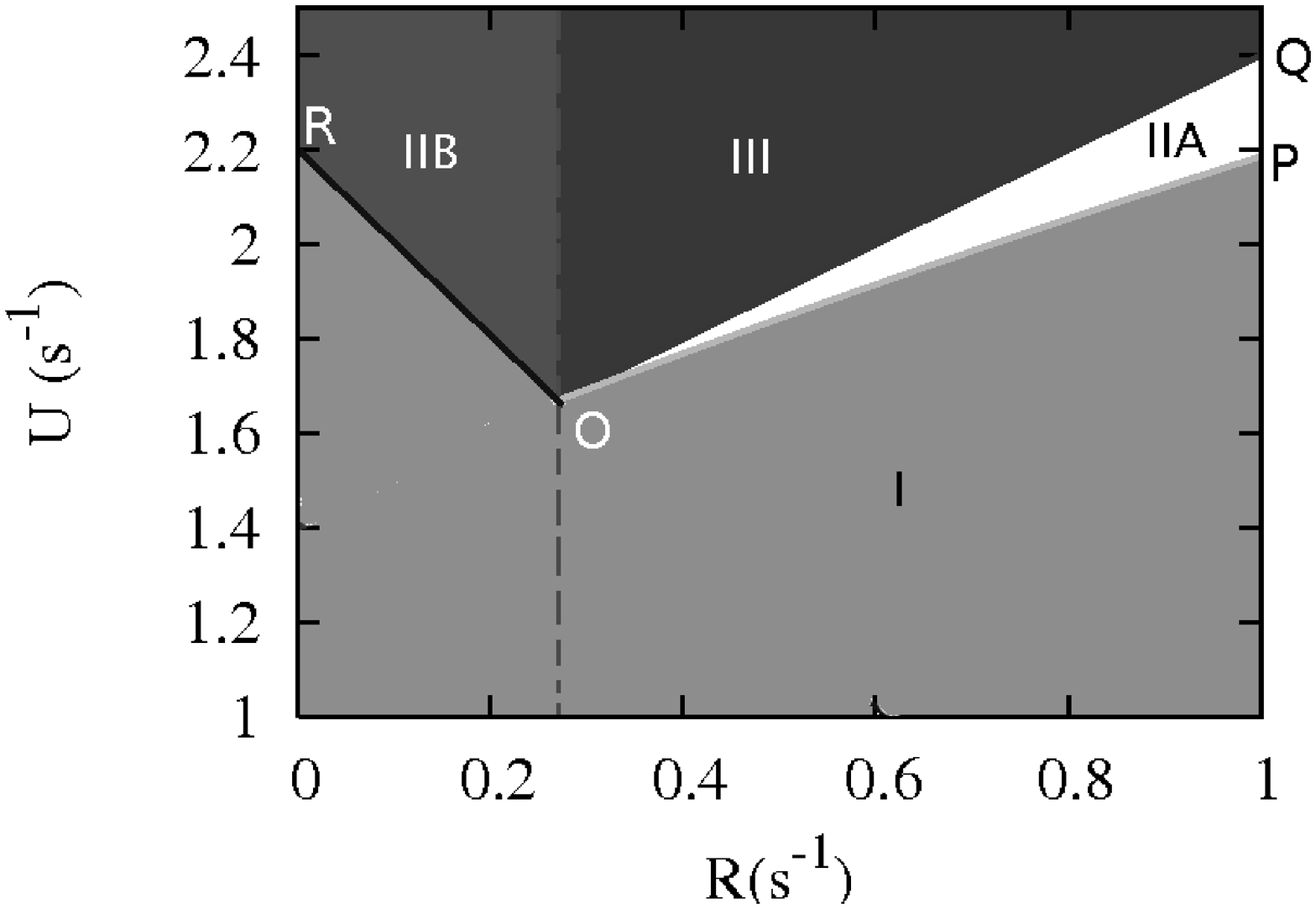}
\caption{\label{phases}}
\end{figure}

 \subsection*{Figure~\ref{phases} }Theoretical phase diagram for the vectorial model
with two ends  in the
variables hydrolysis rate $R$ and on-rate $U$. The line OQ is obtained by setting the cap
velocity equal to zero, and the line OP
is given by the condition $v_{IIA}=0$ where $v_{IIA}$ is the velocity in phase
IIA calculated in Eq.~\ref{vIIA}. Similarly, the line OR is given by the condition $v_{IIB}=0$, where $v_{IIB}$
is the velocity in phase IIB given in Eq.~\ref{vIIB}.

\begin{figure}
\includegraphics[scale=0.4]{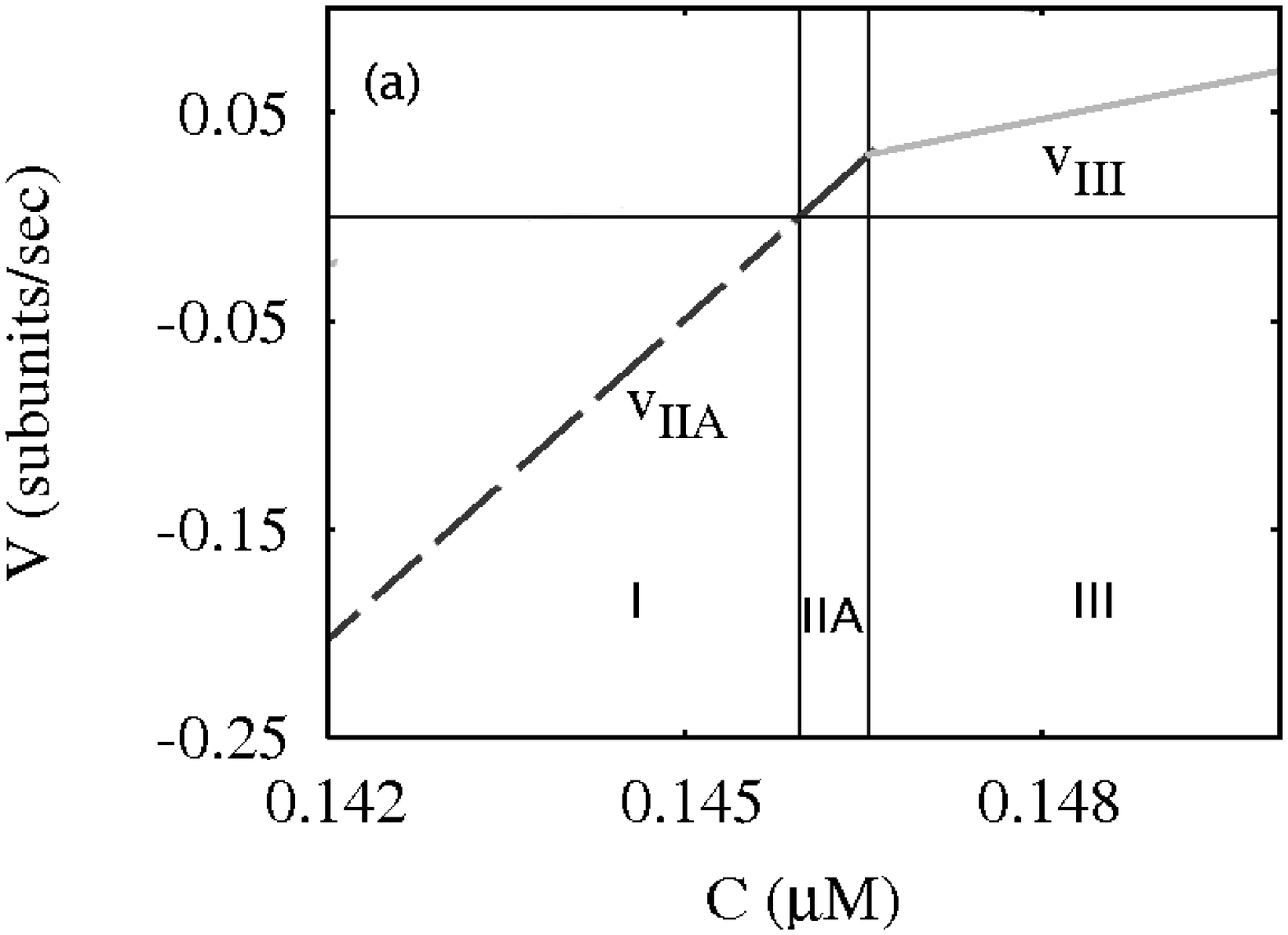}
\includegraphics[scale=0.4]{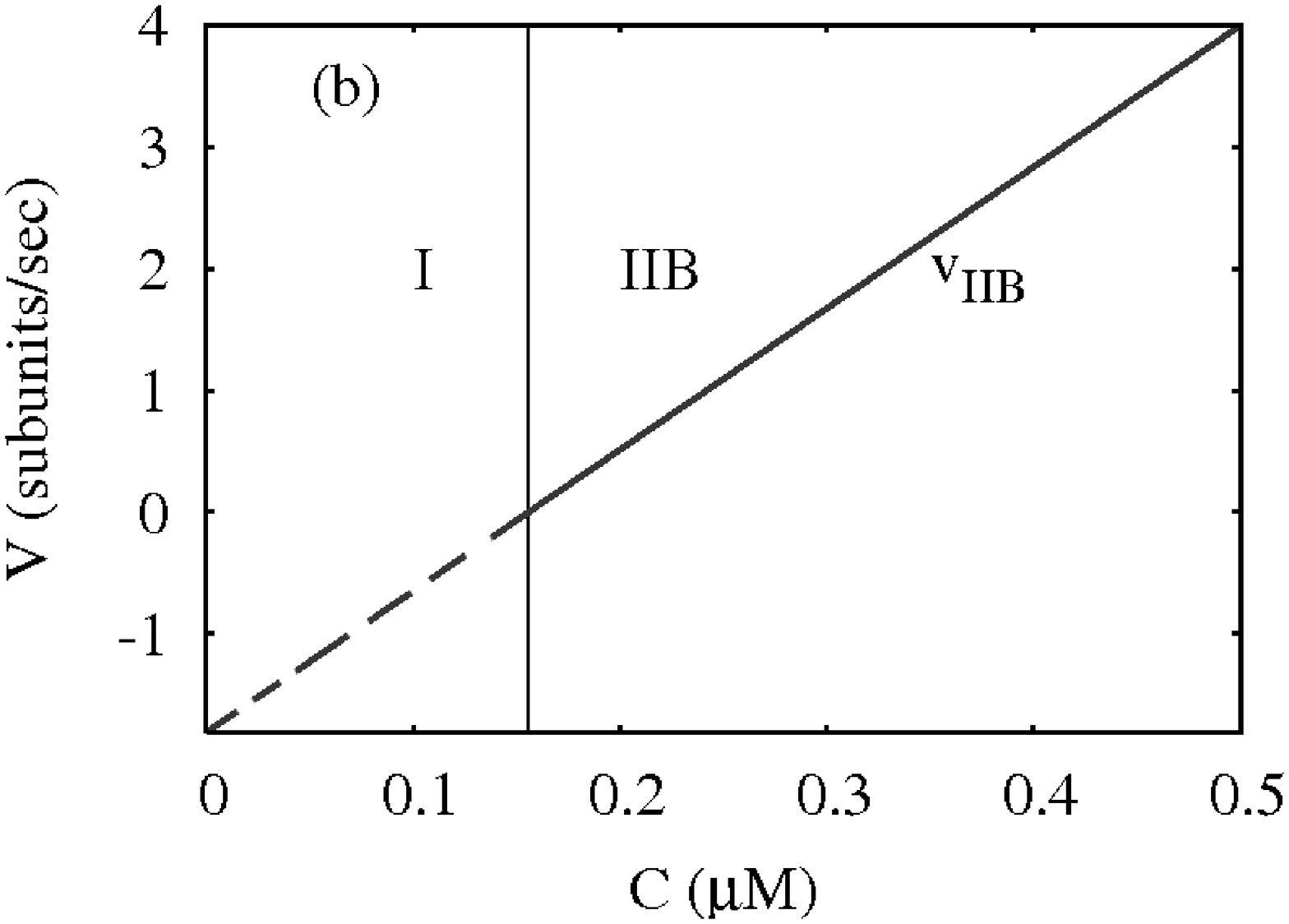}
\caption{\label{fig-velocity}}
\end{figure}

\subsection*{Figure~\ref{fig-velocity}} Filament velocity $v$ versus concentration of free monomers $C$
for the vectorial model with two active ends. (a) Case $R>W_D^-$ for $R=0.3$. In regions I and IIA,
$v=v_{IIA}$, where $v_{IIA}$ is given by Eq.~\ref{vIIA}. In region III, $v=v_{III}$, where the
velocity is that of Eq.~\ref{velocity vIII}.
(b) Case $R<W_D^-$ for $R=0.2$. Here $v=v_{IIB}$ where $v_{IIB}$ is given by Eq.~\ref{vIIB}.

\subsection*{Figure~\ref{fig-avl-bothends}} Average length as function of concentration.
(filled circles ) $W_T^-=0.8$ and $W_D^-=0.27$; (open circles)$W_T^-=0$ and $W_D^-=0.27$; (open squares) $W_T^-=0.8 $
and $W_D^-=0$; (filled squares) $W_T^-=0$ and $W_D^-=0$. The rates which are not specified here
are given in table \ref{table-rates}.  The black line is $ D_{IIA}(c=c_A)/v_{IIA}$

\subsection*{Figure~\ref{fig:phases}} Phase diagram of the random hydrolysis in the coordinate on-rate $U$ versus
hydrolysis rate $r$ (per site).
The symbols have been obtained from Monte Carlo simulations, while
the solid line is the mean-field theory of appendix B. For $r=0$, we recover the
value of $U$ corresponding to the critical concentration of the vectorial model.

\subsection*{Figure~\ref{fig:velocity plots with MFC data}} Velocity versus
free monomer concentration. The squares symbols are experimental data of~\cite{Carlier-etal:86}, which
were taken from Ref.~\cite{carlsson-pb:08}, the solid lines is the velocity for the random model as calculated
from the theory presented in appendix B and the plus symbols is the velocity for the vectorial model using rates in table \ref{table-rates} except for $R=0.12 s^{-1}$ and $W_D^{+}=6.7 s^{-1}$.

\subsection*{Figure~\ref{cdiff}} Diffusion coefficient as function of the monomer concentration
for the random and vectorial model of hydrolysis. The data points are
 the prediction for the random model of hydrolysis while the solid lines are the predictions for the
 vectorial model. The dashed (resp. dash-dotted) vertical line represents the critical concentration
 for the vectorial (resp. random) model.

\begin{figure}
\includegraphics[scale=0.5]{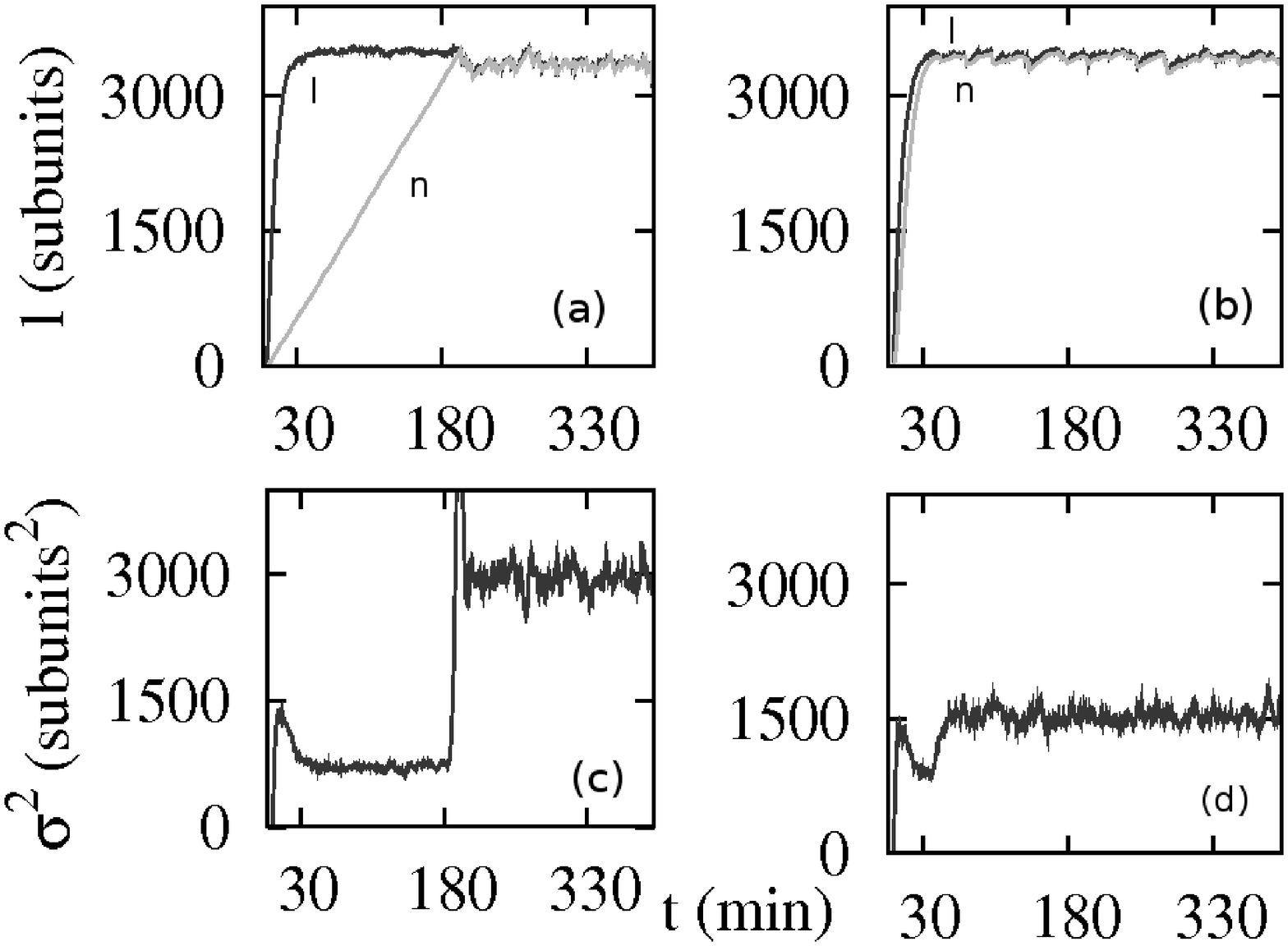}
\caption{\label{Lt-fixedC1}}
\end{figure}

\subsection*{Figure~\ref{Lt-fixedC1}}
(a) and (b) : Total filament length (denoted $l$, black), and total
amount of hydrolyzed subunits (denoted $n$, grey) as function of time
for the case of vectorial hydrolysis (left panel) and random
hydrolysis (right panel) (the total concentration of subunits
$c_T=0.7 \mu M$; 1 filament in a volume of 10 $(\mu m)^3$). Note
that the point where the two curves meet in the random hydrolysis
model occurs much earlier compared to the case of vectorial
hydrolysis ($\approx$10000s). (c) and (d): The variance
($\sigma^2=\langle l^2 \rangle- \langle l \rangle^2$) as a
function of time is plotted  for the vectorial model and  random
model respectively.

\clearpage

\end{document}